# Towards a Taxonomy for Autonomy in Large-Scale Agile Software Development[1]

Casper Lassenius*[†], Torgeir Dingsøyr*[‡]

*Simula Metropolitan Center for Digital Engineering Oslo, Norway

[†]Aalto University, Espoo, Finland

[‡]Norwegian University of Science and Technology Trondheim, Norway

## Abstract

Agile development relies on self-organizing teams having a high degree of autonomy. For single-team develop- ment, more autonomy is generally considered better. In large-scale agile development, where several teams collaborate on the same software with technical and social dependencies, finding the right balance between autonomy and organizational control becomes critical. To this end, it is helpful to have a framework that helps reason about autonomy to help map out to what degree teams can be autonomous, and in what aspects that autonomy needs to be restricted. This paper presents our work towards building a framework for autonomy in large-scale agile development. The preliminary version identifies five levels, and 21 categories of autonomy grouped into three areas. These aspects of autonomy can be taken into account when analyzing or defining the limits of autonomy in large-scale agile software development. The use of the framework is illustrated with an example.

## I. INTRODUCTION

Agile software development relies on self-organizing teams as their basic organizational building block. One of the agile principles states that *The best architectures, requirements, and designs emerge from self-organizing teams* [1].

While that claim certainly is debatable, the idea of self-organization and self-management is either explicit or

---

implied in many agile methodologies. The Scrum guide states that *Scrum teams are...self-managing, meaning they internally decide who does what, when, and how* [2]. Lean software development states that *empowering people, encouraging teamwork, and moving decision-making to the lowest possible level are fundamental to any lean implementation* [3]. In XP, programmers both self-assign to tasks and select with whom they want to work [4]. Teams that are self-organized and self-managed can also be referred to as autonomous teams.

Team autonomy typically refers to the degree of freedom given to individuals or teams and has been widely studied, particularly outside the field software engineering [5], [6].

For small-scale development, having autonomous teams working closely with the customer is widely perceived as an ideal situation, as this brings "decision-making authority to the level of operational problems and uncertainties and thus increases the speed and accuracy of problem-solving." [7].

However, when agile software development is scaled to include several interdependent teams, the issue becomes more difficult: a balance has to be struck between what teams actually can decide themselves and what the organization imposes — a working "sweet spot" between team autonomy and organizational control needs to be identified [8]. Too much autonomy can lead to poor coordination, control, and lower quality [9]. On the other hand, completely depriving teams of autonomy contradicts agile development principles and can lead to lower motivation, less creativity, and poorer team performance.

In the context of large-scale agile development, one approach is to view autonomy and alignment as two separate but interrelated dimensions [10]. However, while intuitively tantalizing, the framework leaves it quite unclear in what dimensions the teams can be autonomous and what facets the goals constrain, making it difficult to apply in practice. Other well-known frameworks, such as the LeSS [11] and SAFe [12] focus on team autonomy as well, with LeSS restricting autonomy significantly less than the SAFe framework [8].

Attempts at applying existing autonomy frameworks to the field also exist. However, applying those frameworks to the specific context of agile software engineering can be challenging, and they are not suitable as-is. [8].

While much of the autonomy discussion in agile software engineering centers around *task autonomy* [13], many other kinds of autonomy can be relevant in the contemporary software engineering context, such as the freedom to choose what process or method and tools to use [14], when to work [15], where to work [16], and with whom to collaborate [17]. This broader and more detailed view to autonomy is not covered by existing scaling frameworks.

Discussing autonomy in software engineering is difficult since we lack a good vocabulary — a taxonomy identifying the main dimensions of autonomy, decision categories, and agents involved.

A further challenge worth noting is that the allocation of decision-making authority might not be static. For example, one company reportedly used the scope of the decision, not its type, to determine who can make it, leading to a situation with contextually determined autonomy [18].

In this paper, we present our preliminary work towards developing a taxonomy for autonomy in large-scale agile software development. This taxonomy can serve a number of purposes: first, the taxonomy gives a more precise definition of autonomy for the specific context of large-scale agile development through a focus on what decisions are made and at which organizational level the decisions are made. Second, this more precise understanding of autonomy can help researchers give a more accurate description of the context of large-scale agile development and could also give practitioners a framework to understand their actual autonomy, which might differ much from perceived autonomy.

## II. BACKGROUND

Large-scale agile development is challenging and the body of knowledge is rapidly growing [19], [20]. However, comparatively little work has been done on autonomy, despite it being a central and somewhat controversial tenet.

### A. Autonomy

Self-organization is a key principle of agile methods in the practitioner literature and has also been studied both in small [7], [21] and large-scale [8], [22], [23] development. Spotify is a well-known example, seeking to *"provide autonomy at every level by allowing employees to make more timely decisions, thereby accelerating teams' innovation performance"* [6]. The Encyclopedia Britannica defines autonomy as *"state or condition of self-governance, or leading one's life according to reasons, values, or desires that are authentically one's own"* [24]. In the research literature, autonomy is often used to describe the degree of freedom given to individuals, teams or projects. Moe et al. [8] use the definition taken from organization psychology [25] defining autonomous work groups as *"employees that typically perform highly related or interdependent jobs, who are identified and identifiable as social unit in an organization, and who are given significant authority and responsibility for many aspects of their work, such as planning, scheduling and assigning tasks to members, and making decisions with economic consequence."*

It is then interesting to understand more about the content and variety of the authority and responsibility at different organizational levels and over various work aspects.

Several studies on agile software development, such as [22] and [8] have used Hackmans authority matrix [5] to distinguish between different levels of autonomy. This taxonomy has four levels, from giving autonomy on decisions regarding i) "executing the task", ii) "monitoring and managing work processes and progress", iii) "designing the performing unit and its organizational context" to iv) "setting the overall direction".

While many studies of agile development take autonomy for granted, a study of a game development team [9] found that "substantial decisions related to work effort, targets, resource allocation and the selection of team members were imposed on them externally". The autonomy was limited *"within a context wherein governing and surveillance mechanisms were defined and activated by actors or authorities outside the team"*. In small-scale agile development, Moe et al. [7] identified barriers to autonomy at the team level (individual commitment, individual leadership, and failure to learn) on the organizational level (shared resources, organizational control, and specialist culture). Further, in large-scale agile development, other identified barriers are related to all four of Hackman´s authority matrix. For example, team task execution (i) is constrained by technology choices in the organization, architectural constraints, and technical debt. In management science, [6] cites works that argue that granting complete autonomy can "give rise to coordination problems, distracting from the company's strategic direction," which can have large consequences in large-scale development.

### B. Areas of Autonomy

As noted in the introduction, most of the discussion in software engineering related to autonomy discusses it at the team level, with a focus on *task autonomy* [13]. However, in software engineering, we think many other issues are relevant, e.g., the freedom to choose what process or method and tools to use [14], designing your own work context,e.g., when to work (deciding your working hours [15], where to work (e.g., in the context of hybrid work) [16], with whom to collaborate [17], and your team composition. Take together, all of these aspects are related to how work is designed.

In practical software engineering work, we think it can be helpful to analyze autonomy also from the point of view of the specifics of software engineering work. While there is abundant literature on software engineering and related activities, we decided to use the most recent version of the software engineering body of knowledge [26] to help us structure the field. The most current version lists 17 knowledge areas, from which we selected the core areas of requirements, architecture, design, construction, testing, operations, and maintenance. In addition, we decided to include the areas software configuration management, models and methods, software engineering management, process, quality and security in our framework. Since we did not see a clear linkage to autonomy for the areas of professional practice or software engineering economics, we decided to leave them out from our preliminary framework.

## III. Method

Most taxonomies have been developed *ad hoc*, but we now have guidelines from both software engineering [27], [28] and information systems [29] on taxonomy development. Here, we chose to present our work according to the widely used method suggested in information systems [29], which focuses on three steps: 1) identifying a meta-characteristic, 2) defining ending criteria, and 3) iterate until the ending criteria are met.

As for a meta characteristic, we wanted to develop a taxonomy of autonomy which shows key decisions and key actors making descions in the context of large-scale agile developemnt.

For ending criteria, we used a mixture of objective and subjective criteria [29]. The objective criteria were examination of prior work on autonomy in software engineering, information systems and management science to identify key dimensions. The key findings from our literature review are reported in the background sections. We also used the subjective criteria suggested by [29]: conciseness, robustness, comprehensiveness, extendibility and explanatory power.

Our approach was a combination of what [29] describes as conceptual-to-empirical and empirical-to-conceptual taxonomy development. We developed the taxonomy with first iteration as part of an interview guide for an ongoing multiple case study, and a second iteration with input from from the literature review.

The first iteration drew heavily on published case studies of large-scale agile development [30], [31] and agile transformation [18], [32] as well as preliminary results from an ongoing multiple case study of team organisation in large-scale agile development. These studies gave us insight both into key roles involved in decision-making and organisation of large-scale agile development.

In the second iteration, we used the studies identified in the literature and in particular the SWEBOK [26] in revising the taxonomy. This specifically led to extending the decision categories from our initial draft.

## IV. Results

### A. Overview of the Taxonomy

An overview of our taxonomy is shown in Table I. The taxonomy identifies three main areas of autonomy: technical core, technical supporting, and work organization and management. Further, it identifies five levels of autonomy: individual, team, leader/expert role, organization, and project. We describe the decision-making areas and the levels in the next subsections.

TABLE I
OVERVIEW OF THE PRELIMINARY TAXONOMY

| Area of Autonomy | Autonomy category | Individual | Team | Leader / Expert Role | Organization | Project |
|---|---|---|---|---|---|---|
| **Technical core** | Requirements | | | | | |
| | Architecture | | | | | |
| | Design | | | | | |
| | Construction | | | | | |
| | Testing | | | | | |
| | Operations | | | | | |
| | Maintenance | | | | | |
| **Technical supporting** | Software configuration mgmt. | | | | | |
| | Quality | | | | | |
| | Security | | | | | |
| **Work organization and management** | Process | | | | | |
| | Models and methods | | | | | |
| | Infrastructure and tools | | | | | |
| | Budgeting | | | | | |
| | Resource allocation | | | | | |
| | Scheduling | | | | | |
| | Measurement and monitoring | | | | | |
| | Work location | | | | | |
| | Work hours | | | | | |
| | Team composition | | | | | |
| | Collaborators | | | | | |

### B. Levels of autonomy

Our taxonomy identifies five levels of autonomy: individual, team, leader/expert role, and organization. In addition, we recognize the concept of a project, which is a cross-cutting organization to the other levels.

The *individual level* refers to decisions that individuals can make regarding their own work. In large-scale agile development, these are primarily the team members in agile teams, include developers, product owners, and coaches/Scrum masters. In some cases, e.g., in LeSS [11] other roles, such as architects or UI designers, can be included in the teams.

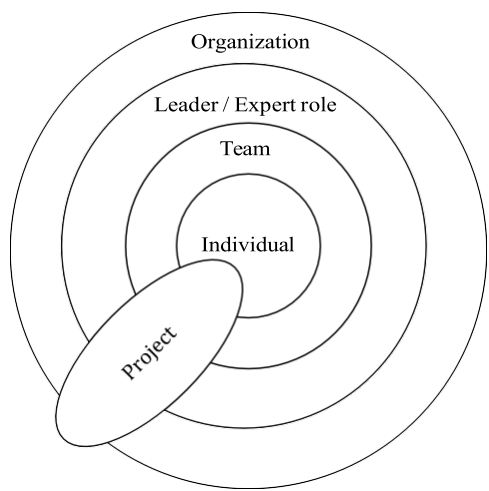

*Fig. 1. Levels of autonomy*

The *team level* refers to decisions made by and inside a single team, as a team. The team can be a normal agile team or have another specified role, e.g., as supporting team. Different team roles are described, e.g., in [33].

The *leader/expert role* refers to decisions made by a team lead, manager, or expert. Expert roles are external to the team and make decisions that affect it. Examples include (chief) architects, UI designers, and domain experts, e.g., legal. In the particular context of the SAFe model [12], this category would also include, e.g., the release train engineer.

The *organization* refers to the larger organization above the team level, including, but not limited to, e.g. Communities of Practice [11], [18, e.g.], HR and other support functions, as well as higher-level line management.

While not included in agile development per se, we decided to include a *project* dimension, as many organizations that have adopted agile development still utilize projects in their development efforts. The project can have roles that have decision-making power, e.g., steering committee, project manager, etc., that need to be considered when planning the decision-making structure. Certain project frameworks, e.g., PRINCE2 [34] have been "adapted" to agile development, and are claimed to be compatible.

### C. Areas of Autonomy

We split the decision areas/types of autonomy into three main categories: technical core, technical supporting, and work organization and management.

The technical core areas represent core areas of software engineering as defined in the SWEBOK. These include requirements, architecture, design, construction, testing, operations, and maintenance [26].

The technical supporting areas include software configuration management, quality, and security [26]. are relevant to work design in modern organizations and that have been discussed in the literature on autonomy, including the aspects of work location [14], working time, team composition [13], and collaborators [13].

### V. Discussion

In this paper, we aimed to understand the key areas and levels of decision-making power for large-scale agile development by proposing a taxonomy of autonomy for this specific context.

Following [27], we argue that taxonomies are important for precise communication and education and also to enable effective reasoning within a research field. Large-scale agile development is an area where there is much jargon from practitioners, which could originate from the need to market certain development methods or intentionally being unclear so that readers can make their own interpretation of messages. For a research field to progress, there needs to be clarity of language, where taxonomies can help.

In the following, we first compare our proposed taxonomy to the one established by Hackman [5] in organizational psychology, which to our knowledge, is the only existing taxonomy. Second, we show how the proposed taxonomy can be used to provide an alternative description of decision-making power in a new analysis of an already published case study. Third, we critically evaluate our taxonomy according to the criteria listed in the method section.

*A. Comparison to existing taxonomy*

The Hackman [5] authority matrix describes four areas of responsibility i) executing tasks, ii) monitoring and managing work processes, iii) designing the performing unit and its context and iv) setting the overall direction. The matrix describes increasing levels of autonomy for units, from manager-led to self-managing to self-designing, and finally, self-governing. A self-governing unit will then have authority on all four levels. In large-scale agile development, one can understand a unit as a team or a larger entity such as a product area.

A problem with the authority matrix as a tool to understand large-scale agile software development is that many decisions are taken at various levels, for example, a team might make decisions on which development process to use at the team level but have to demonstrate work at regular intervals, which are decided at product area level. The proposed taxonomy captures more relevant levels, and also focuses on what we argue are key areas of responsibility in the context of large-scale agile development.

When describing areas of responsibility, why not use an existing taxonomy such as the SWEBOK [26]? The full SWEBOK includes many areas that are not commonly reflected in the decision-making structures in large-scale agile development. Some large-scale agile programmes, such as the Perform programme [30] had a programme structure consisting of four main projects: architecture, business, development and test.

*B. Using the taxonomy*

The taxonomy can be used in several ways: as a way of discussing and determining what kind of autonomy to grant at various levels and as a way of comparing existing cases and frameworks related to large-scale agile development.

The most straightforward way to use the taxonomy is to use it to name the various kinds of autonomy, e.g., testing autonomy, scheduling autonomy, or work location autonomy. At a higher level, one could talk about work design autonomy and technical autonomy. We think using this kind of more specific terms is likely to be more useful than to simply talk about autonomy or autonomous teams, which currently seems to be the norm in agile development.

The taxonomy can be used at various levels of detail: one can select a single area, list major decisions and who can make them under the area, and link them to roles or teams that have the decision-making authority. One could fill in a table like Table I with specifics about roles and their autonomy in each cell, giving a good overview of autonomy in the organization.

Another possibility, which we show an example of in Table II, is to outline, at a high level, at what level the decisions related to an area are made. While obviously masking a lot of detail, even such a simple approach can be enlightening as a means of giving a high-level overview of the autonomy granted at various levels and to various roles.

TABLE II

An Example of using the taxonomy at a high level.

Legends: O=Organization, ER=Expert Role, T=Team, I=Individual, P=Project

| Area of autonomy | Autonomy category | LeSS [11] | SAFe [12] | Org. A | Org. B |
|---|---|---|---|---|---|
| **Technical core** | Requirements | T: PO | T:PO, O:ART | O: PO team | Proj. |
| | Architecture | O: Teams | ER: Architect | O: Teams | ER: Architect |
| | Design | T | ER/T | T | T |
| | Construction | T | T | T | T |
| | Testing | T | T | T+QA Team | T+QA team |
| | Operations | - | - | Customer | Op. dept. |
| | Maintenance | T | T | T | T |
| **Technical supporting** | Software configuration mgmt. | - | - | O | O/T |
| | Quality | - | - | O | O |
| | Security | - | - | O | O |
| **Work organization and management** | Process | O | O | O | T |
| | Models and methods | O/T | O | O | T |
| | Infrastructure and tools | - | O | O | O |
| | Budgeting | - | O | O | O |
| | Resource allocation | - | O | O | O |
| | Scheduling | T | O: ART, T | O: PO team, T | T |
| | Measurement and monitoring | T | O | O | T |
| | Work location | - | - | O/T | T/O |
| | Work hours | - | - | O | O |
| | Team composition | P/O | - | O/T | O/T |
| | Collaborators | - | - | T/I | T/I |

Table II shows a comparison of the LeSS and SAFe scaling frameworks, and a couple of our own previous case studies from the point of view of autonomy using the taxonomy.

In the table, the first letter in the cell represents the level: Organization, Team, Individual, Expert Role, or Project, as shown in the legend. In case there are several levels involved, they are ordered by decision making power with the stronger one (the one with more decision making power) first. In some cases, we have mentioned the specific role or organizational unit responsible; this is done after a colon. For example, in Organization A, testing is decided and conducted on the team level, with support from a QA team.

While by necessity showing little detail, this kind of overview of autonomy can stimulate discussion: it is, e.g., interesting to note that the real team autonomy in both LeSS and SAFe is restricted to mostly the technical areas, except for architecture and requirements that are handled at a higher level. When it comes to work design and supporting activities, the organization and process define most aspects. In contrast, the teams have some autonomy when it comes to task-level scheduling and low-level resource allocation, such as deciding who will work on which task.

In the two exposed cases, we can see that additional autonomy was given to teams and individuals. For example, in Case A, teams were allowed to internally agree on how and when members could work from home (work location), and in both Organization A and B, teams were heavily involved in recruiting and selecting team members (team composition). In Organization B, teams were given additional freedoms, including selecting their own work process, methods, ways of monitoring and reporting, and scheduling. Further, individuals were also allowed to select with whom to work on specific tasks. Initial experiences from Organization B indicate that giving too much autonomy, in particular with respect to process, methods, and monitoring and measurement, can create significant problems in a large organization due to lack of synchronization and predictability.

### C. Limitations

We use the subjective ending conditions and questions proposed by [29], discussing each briefly below.

*Conciseness* A recommendation referred in [29] is to have seven plus/minus two conditions. We have 21 autonomy categories and five organizational levels, a total of 26 dimensions, far above the recommendation and far more than Hackman [5]. However, we have included a nested structure with decision areas and categories which we hope somewhat makes up for the high number of dimensions. Reducing the number would mean a lower precision on decisions or actors.

*Robustness* We believe we have a solid basis in prior work on decision areas, it should be easy to determine differences in autonomy for example in areas such as software architecture and budgeting, and it should be possible to determine which organizational level(s) are involved either relying on descriptions of organizational routines, interviews or observation of practice. There is likely to be differences between what is described in routines and what is actual practice, so this can be seen as a challenge to the robustness of the taxonomy.

*Comprehensiveness* We do not see any projects we know of from practice or literature that we could not classify using the proposed taxonomy. We believe the taxonomy is relevant both for project organizations and product development.

*Extendible* As the taxonomy already has many dimensions, it should not be extended unless it fails to capture important dimensions.

*Explanatory* We argue that it explains key decisions which typically need to be taken in large-scale agile development and who is taking them.

## VI. CONCLUSIONS

We presented a taxonomy designed to help guide decisions about autonomy in large-scale agile software engineering. To our knowledge, this is the first autonomy taxonomy that specifically targets large-scale agile software development. In the future, we plan to validate the framework with practitioners to evaluate its usability, and practical coverage.